# Factors influencing Drug Consumption and Prediction Methods

Denis KOALA, Zakaria YAHOUNI, Gülgün ALPAN, Yannick FREIN

Univ. Grenoble Alpes, CNRS, Grenoble INP[1] , G-SCOP, 38000 Grenoble, France

{denis.koala; zakarya.yahouni; gulgun.alpan; yannick.frein}@grenoble-inp.fr

*Résumé*
Les problèmes de logistique hospitalière, de gestions de stocks et d'estimations des besoins de produits de santé se posent encore de nos jours dans les centres hospitaliers qui doivent conjuguer budgets restreints et satisfactions de patients. De nouvelles perspectives devraient s'imposer pour résoudre les problèmes en amont, surtout la prédiction des besoins. Ce travail se plonge dans la littérature afin de mettre en évidence les méthodes de quantification et d'estimation des besoins des produits de santé dans les établissements de santé. Un second objectif est de dresser une liste des facteurs qui impactent la consommation des médicaments, notamment les facteurs qui sont souvent utilisés dans les méthodes de quantification. Cette revue de la littérature identifie six méthodes utilisées par les praticiens des quatre coins du monde en tenant compte de certains prérequis et de type de données. De même, 34 facteurs sont identifiés et regroupés en trois catégories. Ces résultats devraient permettre de mettre sur pied de nouveaux outils de prédiction des besoins en médicaments, de faciliter en amont le dimensionnement de nouveaux entrepôts pharmaceutiques et de résoudre des problèmes de logistique hospitalière.

*Mots clés* - Consommation de médicament ; facteur ; quantification ; prédiction ; logistique hospitalière.

*Abstract*
Estimating the needs of healthcare products and inventory management are still challenging issues in hospitals nowadays. Centers are supposed to cope with tight budgets and patient satisfaction at the same time. Some issues can be tackled in advance, especially regarding the prediction of drug consumption needs. This work delves into the literature in order to highlight existing methods of quantifying and estimating the needs for drugs in health facilities. A second objective is to draw up a list of factors that impact drug consumption in particular, factors that are used in these prediction methods. Following this literature review, it appears that six sustainable methods are being used by practitioners around the world, taking into account certain prerequisites and types of data. Thirty-four factors are identified as well and grouped into three categories. These results should participate in setting up new tools for predicting the need of drugs, to facilitate the upstream dimensioning of new pharmaceutical warehouses and to solve some hospital logistics issues.

*Keywords:* Drug consumption; factor; quantification; prediction; hospital logistic

## 1 INTRODUCTION

In France, as in other developed countries, the healthcare sector is constantly transforming to adapt to a complex and changing environment with increasingly tight budgets. OECD[2] reports [OECD, 2019; OECD, 2015; OECD, 2013] reveal that over the period 2000-2008, average per capita health expenditure per year increased by 4.1%, 1% from 2008 to 2013 and 2.4% from 2013 to 2018. In such a context, hospital centers are obliged to optimize their management. They have launched several series of investment programs in order to achieve this, while guaranteeing the safety and quality of the care provided to patients. The Hospital Plan from 2007 to 2012 in France [Safon, 2017] is a striking example. This has shown limits, as [Bernardini-Periniciolo et al, 2018] states. Hospital expenditures represent about 40% of healthcare expenditures [OECD, 2013], of which 30 to 46% are related to logistics activities, pharmaceutical logistics in particular [Nachtmann et al, 2009; Poulin, 2003], making hospital logistics the second largest component of expenditure after personnel expenses [Poulin, 2003; Volland et al, 2017]. Pharmaceuticals account for 70-80% of supply costs [Rego and al 2014; Kelle, 2012]. There is therefore an interesting potential for gains.

Compared to other sectors such as industry, hospital logistics has not been considered a priority in the research world in the past. Several reasons have been pointed out such as legal aspects, the high complexity of supply chains, the efficiency of patient treatment, the stochastic and unpredictable nature of product demands as well as the importance of the human factor, etc. [Beier, 1995 ; Almarsdóttir et al, 2005 ; Moons et al 2019 ; Romero, 2013 ; Volland et al, 2017]. The research potential is therefore interesting. Lack of innovation in hospital logistics leads to problematic situations in hospital departments regarding inventory management, unjustified forecasting techniques, lack of IT support, etc. [Rachmania et al 2013; Romero, 2013].

Over the last twenty years, logistics has been identified as a key lever to manage healthcare costs [Dacosta-Claro et al 2002; De Vries, 2011]. Fruitful research has been conducted and methods have been implemented to optimize and curb certain deficiencies, inventory management problems in particular (chronic and generalized shortages, surpluses or overstocking, etc.), supply chain problems, etc. [De Vries, 2011 ; Jurado et al 2016 ; Battersby 1993 ; Kelle et al, 2012 ; Jurado et al 2016 ; Maestre, 2018].

---

[1] Institute of Engineering Univ. Grenoble Alpes
[2] OECD: Organization for Economic Co-operation and Development

Some research estimates that through effective logistics management, approximately half of the costs associated with hospital logistics can be eliminated [Poulin, 2003; Romero, 2013; Volland et al, 2017] although the literature on this issue is still rare [Kharroubi, 2019]. A prerequisite for an efficient stock management is to have a correct dimension of the stock. This is a strategic problem and needs to be decided when the pharmaceutical warehouses are to be built.

In fact, setting up a new logistics warehouse for a hospital pharmacy can take several months, with corrective actions in case of incorrect dimensioning. To ensure the most accurate dimensioning and to guarantee the performance of the pharmacy supply chain within a hospital structure, it would be necessary to implement decision support tools that are both fast and efficient. The dimensioning and inventory management are strongly influenced by the consumption behavior of pharmaceutical products, which in turn is spurred by several factors [Kharroubi, 2019]. This work is positioned in this perspective.

In this work, we highlight the existing literature methods that aim to facilitate the quantification and estimation of drug needs from the 1980s to the present day. This cannot be done without taking into account the causal links and factors that impact consumption. The way in which health product quantification is carried out has serious implications for the health system of a country, a region, even the operation of a health facility. If projected estimates are lower than real needs, the impact can seriously hamper the effective delivery of needed health care: similarly, if the estimated quantities are excessive, the result is a waste of scarce resources. Wasteful or irrational drug use can be perpetuated by simply continuing to order products based on information from the historical use of each drug [WHO, 1988; MSH, 2012]. Significant shortages and overstocking can occur by ordering on the basis of theoretically determined quantities that have not been sufficiently tested. It is obvious that the process of estimating drug needs should be done by considering a certain number of indicators and factors of consumption. Additional complexity arises when estimating drug needs when dealing with exceptional situations such as disasters, epidemics or pandemics like COVID-19. These phenomena are difficult to model and predict. They are considered in certain estimation methods by adjustment [ MSH, 2012]. However, the effectiveness is not very accurate, especially when the availability of health products is not the sole responsibility of hospitals (eg. :  Masks in COVID-19 situation).  In this work we do not consider such extreme and rare situations.

Section 2 and 3 are respectively dedicated to the research methodology, followed by a description of the quantification methods found in the literature with a brief comparison. Section 4 addresses the factors influencing drug consumption. As concluding remarks, a research perspective is issued, regarding hospital and pharmaceutical logistics in particular.

## 2 METHODOLOGY

### 2.1 Identification of publications

To identify relevant literature, we proceeded in several steps. The first step consisted in using the google scholar and direct science platforms to identify appropriate keywords for successful searches. Based on that, three classes of keywords were drawn up as shown in Table 1. The first-class groups together terminologies related to the quantitative and predictive aspect of needs, while the second class is an addition to the first class with a view to contextualizing and centralizing research. Depending on linguistic affinity, terms such as "drug" or "medicine" can be found in the literature but are terms to designate the same entity. The last class of keywords consists in highlighting consumption indicators and the work that is carried out in the context of stock sizing and management. A second search on the same databases was conducted for relevant keywords, which defined our second step. Obviously at this stage some keywords were excluded like "hospital", "influence", "stock inventory", "stock capacity" which for some are too broad and others a bad combination. In the third step a cross-referencing of the keywords in the platforms, google scholar, direct science, web of science and IEEE was carried out. Although the results were more or less satisfactory, we used some databases such as WHO[3] and MSH[4]. The search is limited to documents in English and French over the period 1980 to 2019. Although we initially focused solely on articles, the request for official reports proved to be a plus, particularly on quantification methods.

Once the papers were grouped together, the fourth step was to eliminate duplicates. Finally, in the last step we eliminated papers that do not address our problem. These different steps resulted in 81 papers related to quantification methods and 59 related to drug consumption factors.

| **Class 1** | **Class 2** | **Class 3** |
|---|---|---|
| "Estimating" | "Drug utilization" | "Inventory management" |
| "Forecasting" | "Essential drugs" | "Stock Inventory" |
| "Quantification" | "Pharmaceutical requirements" | "Hospital logistic" |
| "Prediction" | "Medicine consumption" | "Stock capacity" |
| "Quantifying" | "Drug consumption" | "Capacity planning" |
| "Forecast demand" | "Hospital pharmacy" | "Warehouse management" |
| | "Hospital" | "Influence" |
| | | "Factor" |

**Table 1: Classes of Keywords**

### 2.2 Scope

As a reminder, pharmaceuticals include items directly related to patient care, essential drugs, medical consumables, labile blood products, food, linen, sterile items, etc. In France, healthcare products are grouped into 20 groups[5]. The paper focuses on

---

[3] WHO: World Health Organization
https://www.who.int/home/search

[4] MSH: Management Sciences for Health
https://www.msh.org/search

[5] French Website of health products categories
https://www.legifrance.gouv.fr/affichCodeArticle.do;jsessionid=8569371CC949E72C401CFBC44FDFFF32.tpdila16v_3?idArticle=LEGIARTI000033897163&cidTexte=LEGITEXT000006072665&categorieLien=id&dateText

essential drugs and the factors that influence their consumption in hospitals.

## 3 METHODS FOR DRUG NEEDS QUANTIFICATION

The effective introduction of an action plan for the consumption of essential drugs and vaccines requires a good supply management system. This system is composed of selection, quantification, procurement, distribution and use [WHO, 1988; MSH, 2012]. The selection consists of determining which essential drugs are needed. It is a process that begins by defining a list of common diseases for each level of health care. It has a considerable impact on the quality of care and the cost of treatment, so it is one of the pillars of an effective health system. Quantification is the process of estimating the quantities and costs of drugs and health products needed for a specific time period and determining when shipments of these products must be delivered to ensure an optimal and uninterrupted supply.

Procurement is the process of choosing suppliers, placing and monitoring orders, controlling the quantity and quality of medicines and paying suppliers. Then comes distribution, which includes reception, storage, inventory, transport and record keeping, i.e. information collection and control. Prescribing, dispensing and use of medicines; patient compliance concerns the use phase of the system.

In the context of our research, we are interested in the quantification process. Even though some of the methods that result from it include selection, based on the principle that the drugs must first be determined by the health institution or organization concerned.

Needs are estimated according to a given context, and the analysis must include contextual factors. It should be noted that, according to WHO, the effectiveness of quantification depends on the quality of available information and resources and accurate data on morbidity and drug use.

According to [MSH, 2012], quantification is not a simple calculation, it is the first step in the procurement process. There are many aspects to quantification such as: Calculation of estimated order quantities, costs and delivery dates of shipments; Planning, mobilizing and securing financial resources; Estimation of storage requirements; Assessing the rational use of products; Facilitate coordination of procurement and logistics with donors, suppliers, health facilities and other stakeholders; Inform manufacturers and suppliers about future demand for manufacturing, procurement and logistics management decisions, etc.

The quantification process is normally applied for: calculation of quantities ordered for public procurement, estimate budgetary requirements, develop purchasing quantities for new programs, expand procurement quantities for scale-up programs, etc. (see [WHO,1988] and [MSH, 2012] for more details).

Since common sense does not allow us to deal with certain complexities in quantification, methods have been developed since the 1980s, the identified ones are:

- Consumption method
- Morbidity method
- Proxy consumption methods
- Service-level projection of budget requirement
- Hospital Request method
- Population-based method

The methods most encountered in the literature are the first three in the list [Soeters, 1988; WHO, 1988; Osore, 1989; MSH, 2012]. Drugs could be quantified by using one or a combination of these approaches. Those methods are normally used to quantify needs for an annual or semiannual procurement. We will briefly explain each one of them in the following paragraphs.

### 3.1 Consumption method

This method is based on a facility's existing drug consumption. In fact, a list of all drugs eligible to be ordered or purchased is prepared and a more accurate inventory of past consumption, usually over a recent period of six to twelve months, is used to calculate the quantities required for each drug for each facility. An analysis of facility consumption is necessary and whenever it appears abnormal for a given drug, a correction is made by adjusting it upwards or downwards until it reaches an adequate level. The adjustment is already made according to stock shortages in order to obtain the average monthly consumption. Then, the average monthly consumption is multiplied by the number of months to be covered by the purchase. Safety stock levels and delivery times (in months) are also multiplied by the projected monthly consumption. These three figures are added together to obtain the gross requirements for the period, subtracting the usable stock on hand and any stock on order from the gross estimate, to deduct the quantity to be purchased. A further adjustment is then made for losses. And if cost is a factor in the quantification then the expected unit cost for each is multiplied by the number of units to be purchased to obtain the expected purchase value for the total quantity which is an addition of the purchase values of each drug on the list. Finally, a final adjustment is made if this cost is higher than the budget. This is why the method is also called the adjusted consumption method [WHO, 1988]. In the 1980s, WHO proposed three main steps to take into account the approach to quantify the drugs to be made available over the forecast period. In 2012, the MSH group developed 11 steps that included estimating the total cost of forecasting with the Quantimed tool (see [MSH, 2012] for more details).

In order to use this approach, preconditions must be met [WHO, 1988]. Accurate and reliable consumption data that are available or can be obtained relatively easily should be available. The same applies to essential drug lists with their packaging and prices. Losses due to expiration, damage and theft should not be excessive. Also, the supply of medicines in "typical" facilities should be enough (in practice, there has been no shortage of essential medicines for more than three months of the year). There should also be a record of suppliers' delivery times and budgets to be allocated.

While this method of quantification is inadequate for new hospitals, it is generally preferable for stable programs where funding, pharmaceutical management and prescribing are reasonably satisfactory. It is also easier to use in environments such as hospitals with many health problems and complex treatments.

### 3.2 Morbidity method

Based on the number of people suffering from a given disease at a given time in a given population, according to prevalence or incidence, the so-called morbidity method was established. This method uses data on patient health facility attendance and morbidity to predict the need for drugs based on assumptions about how problems will be treated. In this sense, it requires data sets: a list of common health problems, a list of essential medicines that includes therapy for the problems, and a standard treatment package for quantification either on the basis of standard treatment regimens agreed upon for each defined health problem or on the basis of current practice.

Although very demanding in terms of data, this method is generally preferable for new or rapidly changing services or when services are radically reorganized. It is also preferable if prescribing practices are costly and irrational, as it provides a systematic basis for improvement.

To carry out its application and to obtain the most accurate estimate of the quantity of products and their costs, [MSH, 2012] presents this method with 12 steps in the process from specifying lists to comparing costs and budget to making corrections.

*3.3    Proxy consumption method*

Previous approaches require data that is often very precise, and some systems are faced with information deficits. In this way they are not feasible, and the proxy consumption method presents itself as an alternative. In this method, consumption data are extrapolated from a database in another health system or another region with similarities. For this reason, it is used for new facilities that may lack data.

This method can be based on two approaches. The first population-based approach is to define drug use per 1,000 inhabitants and the second service-based approach is to establish drug use by specific cases or by inpatient admissions of patients. It is also advisable, for a complete quantification, to use a combination of the two approaches.

For its application, 9 steps are established for successful quantification, from the selection of the standard system for comparison and extrapolation, the definition of product lists to the estimation of the costs of each and total drug [MSH, 2012]. It is a method that is quite concerned with the indicators and factors of consumption. Also, standard facilities should closely resemble the region or country for which the estimate is made as a main condition to get a good prediction. The resemblance should be in terms of geography and climate, patient population served, morbidity patterns, prescribing practices, standard treatment guidelines, essential medicines lists and pharmaceutical supply status.

*3.4    Service-level projection of budget requirement*

The approach outlined here is generally used to estimate financial needs, not specific quantities of drugs, for the purchase of pharmaceuticals on the basis of costs per patient treated at different levels of the same health system or on the basis of data from other health systems.

Like the proxy method of consumption, this method is an extrapolation method that leads to rough estimates given the large variations that may exist between the target health system and the system used as the standard data source. Sources of error encountered include: prescribers in the target system using a different drug combination than in the source system, variability in disease frequency and patient visits per facility, and differences in the effectiveness of procurement and financial management systems in the two settings. It does not anticipate specific drug needs but provides a clear and logical rationale for pharmaceutical funding requirements [MSH, 2012]. For its application, the main requirement of this method is a fairly reliable estimate of the average cost of medicines per patient and the average number of visits at different levels of the standard health system. The data that must be compiled are: the average number of curative and non-curative outpatient visits and bed days and/or other types of patient contact for each type of facility in the original health care system; and the average cost per curative and non-curative outpatient visit and bed day and/or other type of patient contact in each type of facility in the original health care system.

*3.5    Hospital request method*

This method is considered one of the oldest approaches for quantifying and estimating needs. It is a method of calculating drug requirements that is based on hospital requests for national, regional or warehouse pharmacies. The free choice of drugs to individual prescribers could be mentioned as an advantage of this method [Soeters, 1988; WHO, 2014]. We can also add the reduction of planning needs and guarantee a set of products will be available but are limited in the type and quantity of medicines and lead to significant waste and stock-outs. However, problems may include: hospital requests for more drugs than necessary, inadequate hospital requests (taking into account the cost/therapeutic ratio) or unnecessary drugs through personal preferences or the influence of pharmaceutical companies, health care staff often do less than optimal prescribing practices (polypharmacy, over-prescribing and unnecessary prescribing) [Soeters, 1988].

*3.6    Population-based method*

The last method that is elucidated in this paper is the population-based method which is based on the prevalence of various diseases in the population. The population-based method depends on total need that is different from demand. Information that is needed includes demographic and disease monitoring data e.g. number of new patients, number of continuing patients, those needing to change to a different treatment, in addition to disease prevalence rates among the target population. A formula is required for each product which can be unrealistic for medicines with multiple uses [WHO, 2014]. However, as can be expected, this theoretically interesting method is not used in practice because the required data are very difficult to obtain, implying very few references in the literature [Soeters, 1988]. Among the data we can note socio-economic data such as age, sex, marital status, co-morbidity, geographical position of patients. These difficulties were even more real in the first years of application of the approach because computer tools were lacking in health structures and data tools were not yet on the policy agenda.

*3.7    Comparison of methods*

Based on these findings, a comparison has been made in this document by highlighting the advantages and limitations of each method summarized in Table 2.

Among the six methods highlighted in the literature, the most widely used are the consumption, morbidity and proxy consumption methods. The reasons that can be put forward are the availability and specificity of the data that fall within the scope of the application of quantification and the factors that can have an impact on consumption in health facilities.

The method considered in the literature to be the most accurate for quantifying the use of pharmaceuticals is the consumption-based approach. To apply this method, data should be complete, accurate, and correctly adjusted for out-of-stock periods and anticipated changes in demand and use. However, the main problem with this method is that it does not normally consider the relevance of past consumption patterns, which may or may not correspond to public health priorities and needs regarding morbidity. This can lead to irrational drug use. If stock-outs have been widespread for long periods of time, it may not be possible to apply this method accurately, which is why capturing actual demand is essential for the most accurate approach.

Quantification based on morbidity is the most complex and time-consuming method to apply, according to what is said in the literature and the type of data input. Indeed, in many countries, it is very difficult to collect valid morbidity data for several diseases; therefore, some needs will be neglected in quantification. Data on patient consultations are often incomplete and inaccurate, and it is difficult to predict what percentage of prescribers will actually follow the standard treatment regimens used for quantification. Despite these constraints, this method may still be the best alternative for procurement planning or for estimating budgetary needs in a system or health facility. The method is also often used to estimate the cost of drugs in a large number of settings, such as a small primary care system, a specialty hospital, or a new hospital.

| | Advantages | Disadvantages |
|---|---|---|
| *Consumption method* | <ul><li>Detailed morbidity data and standard treatment regimens are not required.</li><li>It requires less detailed calculations.</li><li>It is useful for institutions such as hospitals, where there are many problems and complex drug treatments.</li><li>It is reliable if consumption is well recorded and stable and is not likely to differ significantly from the current supply.</li><li>It identifies inventory management problems and encourages improvements.</li><li>Easy method, requiring only one study by a few people.</li><li>Improves the availability of essential medicines and improves cost-effectiveness</li></ul> | <ul><li>Must have accurate consumption data.</li><li>Reliable data on pharmaceutical consumption can be difficult to obtain, especially in new or rapidly changing services.</li><li>It does not provide a detailed or systematic basis for reviewing drug use and improving prescribing; if the prescribing pattern is unsatisfactory and is not corrected, this method risks perpetuating it.</li><li>It is unreliable if there have been long stock-outs (more than three months) or significant loss or waste of medicines.</li><li>It is not conducive to good morbidity recording.</li><li>Does not interact with the existing information system Previous drug shortages can lead to biased results.</li><li>Existing (non-optimal) prescribing practices are accepted.</li><li>Changes in demand and use including irrational use of medicines</li></ul> |
| *Morbidity method* | <ul><li>No need for pharmaceutical consumption data; the method can be used for new services that do not have this data.</li><li>Based on a rational prescribing system, it provides a systematic basis for reviewing drug use and prescribing, particularly at the primary care level where drug treatments are less frequent and simpler.</li><li>Improves prescribing habits and interacts with the existing information system. Improves the availability of essential drugs, reduces the consumption of non-essential drugs and improves cost-effectiveness.</li><li>Promotes reliable recording of morbidity.</li></ul> | <ul><li>Problems can arise with detailed morbidity data and agreed-upon standard treatment regimens.</li><li>It requires more detailed calculations.</li><li>Results may differ significantly from the actual drug supply.</li><li>Supply will not match utilization if standard treatments are not followed.</li><li>Assesses only the quantities needed to treat patients; losses and wastage must be considered separately.</li><li>Exact hospitalizations are difficult to predict</li></ul> |
| *Proxy consumption method* | <ul><li>Not demanding in terms of the amount of data on morbidity and pharmaceutical consumption</li></ul> | <ul><li>Comparability of facilities, morbidity patterns and treatment practices between sites/countries,</li><li>Incomplete or inaccurate consumption data from the reference facility.</li></ul> |
| *Service-level projection of budget requirement* | <ul><li>Reliable estimate of average medicine cost per patient</li><li>No need for consumption details and morbidity data</li></ul> | <ul><li>Does not forecast needs for specific medicines,</li><li>Variable facility use,</li><li>Attendance treatment patterns,</li><li>Supply system efficiency.</li></ul> |
| *Hospital Request method* | <ul><li>The free choice of drugs by prescribers, especially for centralized health systems (countries, regions, etc.).</li></ul> | <ul><li>Hospitals are requesting more medication than necessary, knowing that the quantities will be reduced anyway.</li><li>Hospitals request drugs that are inappropriate (cost/therapeutic) or unnecessary because of personal preferences or the influence of pharmaceutical companies.</li><li>Health care staff often adopt suboptimal prescribing practices</li><li>Expensive, non-essential drugs are required.</li><li>Non-optimal prescribing practices are accepted.</li><li>Prescribers are sensitive to the marketing practices of pharmaceutical companies and consumer pressure.</li></ul> |
| *Population-based method* | <ul><li>Interesting for drugs that do not have multiple uses</li></ul> | <ul><li>Data are very difficult to obtain, especially patient socioeconomic information</li><li>Difficult to quantifying medicines with multiple uses</li></ul> |

**Table 2: Methods advantages and limits**

Proxy consumption is the method generally used when neither the consumption-based nor the morbidity-based method is feasible. This method is more likely to provide accurate projections when used to extrapolate one set of facilities from another set of facilities serving the same type of population in the same type of geographic and climatic environment. If the method is applied using standard data from another country, the results will only be rough estimates. Even when both target and health facilities are similar, it is very difficult to assume that disease incidence, patterns of use, and prescribing habits will be essentially the same in both cases. Nevertheless, this method may be the best alternative in the absence of appropriate data required for the consumption or morbidity-based method. The surrogate consumption method is also useful for cross-checking projections made with other methods All of the methods reported are based on a number of factors and indicators related to drug use. Therefore, a determination of the factors that influence consumption should contribute to both a judicious choice of method and the most accurate quantification.

## 4 FACTORS

| Categories of factors | N° | Factors | Sources |
|---|---|---|---|
| Socio-demographic and Socioeconomic | 1 | Gender (Sex) | [Pappa et al, 2008 ; Jiménez-Rubio et al, 2010 ; Rocío et al, 2018 ; Mishuk et al, 2018 ; Howard et al, 2018 ; Henricson et al, 1998 ; Sarwar et al, 2017 ; Borrell et al, 2010] |
| | 2 | Age | [Pappa et al 2008 ; Jiménez-Rubio et al, 2010 ; Rocío et al, 2018 ; Mishuk et al, 2018 ; Howard et al, 2018 ; May et al, 1974 ; Henricson et al, 1998 ; Pokela et al, 2014] |
| | 3 | Race/ethnicity/country of birth | [Mishuk et al, 2018 ; Howard et al, 2018 ; Henricson et al, 1998] |
| | 4 | Income(annual) | [Jiménez-Rubio et al, 2010 ; Matin et al, 2015 ; Mishuk et al, 2018 ; Howard et al, 2018 ; Henricson et al, 1998] |
| | 5 | Marital/Civil status | [Pappa et al 2008 ; Rocío et al, 2018 ; Sarwar et al, 2017] |
| | 6 | Education | [Pappa et al 2008 ; Jiménez-Rubio,2010 ; Henricson et al, 1998 ; Pokela et al,2014] |
| | 7 | Residence(rural/urban) | [Pappa et al 2008 ; Matin et al, 2015 ; Henricson et al, 1998] |
| | 8 | Employment or Activity status (employed, retired, unemployed, inactive) | [Jiménez-Rubio, 2010 ; Henricson et al, 1998 ; Sarwar et al, 2017] |
| | 9 | Social class | [Rocío et al, 2018 ; Borrell et al, 2010] |
| Health-related | 10 | Health insurance | [Jiménez-Rubio, 2010 ; Howard et al, 2018] |
| | 11 | Health risk or Lifestyle (smoking; alcohol consumption, obesity: Body Mass) | [Rocío et al, 2018] |
| | 12 | Chronic conditions/chronic diseases | [Pappa et al 2008 ; Jiménez-Rubio et al, 2010 ; Howard et al, 2018] |
| | 13 | Comorbidity | [Howard et al, 2018] |
| | 14 | Prior experience with drug use | [Howard et al, 2018] |
| Facility related & staff member | 15 | The drug price index | [Berndt, 2002 ; Pappa et al 2008] |
| | 16 | Number of physicians for consultations and hospitalization | [Pappa et al 2008;] |
| | 17 | The number of hospital beds | [Pappa et al 2008 ; Matin et al, 2015] |
| | 18 | The number of medical visits | [Matin et al, 2015] |
| | 19 | Morbidity patterns (frequency of health problem) | [Matin et al, 2015;] |
| | 20 | Number of doses of each medicine per day | [Muhia et al, 2017 ; Henricson et al, 1998 ; Sarwar et al, 2017] |
| | 21 | Geographic area (region, district) | [WHO, 1988 ; Alkan et al, 2015] |
| | 22 | Facilities geographic position | [WHO, 1988; Alkan et al, 2015] |
| | 23 | Patient contact for each category of facility | [WHO, 1988] |
| | 24 | Service levels (Services (bed- day)) | [Muhia et al, 2017] |
| | 25 | Number and type of health facilities | [Muhia et al, 2017] |
| | 26 | Seasonal factors (and replacing an older medicine) | [ MSH, 2012] |
| | 27 | Periods (lead time, time to review) | [MSH, 2012] |
| | 28 | Medical Specialist (health workers characteristics) | [Mishuk et al, 2018; MSH, 2012] |
| | 29 | Storage point (capacity) | [Julius et al, 2018 ; Rocío et al, 2018] |
| | 30 | Procurement processes | [MSH, 2012] |
| | 31 | Budgeting processes | [Julius et al, 2018] |
| | 32 | Legal requirements | [Julius et al, 2018] |
| | 33 | New programs, expansion of existing services | [Julius et al, 2018] |
| | 34 | Technologies of facilities | [Howard et al, 2018] |

**Table 3: Factors affecting drug consumption in hospitals**

In the literature, there are several variations of factors that are considered in the quantification of pharmaceutical needs. A number of papers [Pappa et al 2008 ; Jiménez-Rubio et al, 2010; Rocío et al, 2018; Mishuk et al, 2018 ; Howard et al, 2018 ; May et al, 1974; Henricson et al, 1998; Pokela et al, 2014] refer to factors related to patients, to the population to be taken into account in the healthcare system, generally socio-economic factors related to the medical system. Various studies have been carried out, some for specific products and others for product categories, taking into account the classification of the 29 WHO categories, in all four corners of the world. To summarize, these factors have been subdivided into three categories and presented in Table 3.

The first category includes socio-economic factors such as: gender, age, origin of the patient, annual income, social class, and so on. Some studies conducted in the USA on the correlation between drug use and race and origin are not applicable in all countries. In the second category comes the factors related to the health of the patient. These include factors such as co-morbidity, health insurance, lifestyle (smoking; alcohol consumption, obesity: Body Mass, etc.).

The last category includes factors related to health care institutions and personnel.

All of these factors could be parameters for estimating drug requirements for a new hospital structure, or for good inventory management. It shall be noted that some factors are not applicable in all health systems, for example, race or country of birth, which are considered to be impacting factors in the United States [Mishuk et al, 2018], may not be significant in another country. Some studies related to consumption factors found in the literature were less generalized. For instance, some factors focus on specific drugs and types of products such as generic drugs [Howard et al, 2018], ARVs, tuberculosis, etc. Also, among the factors in this list some may not be possible to integrate in a prediction model if the related data is not available.

## 5 CONCLUSION AND PERSPECTIVES

Looking forward to the well-being of patients, hospitals are facing several challenges. Among these challenges, reducing the cost of logistics, which represents the second most important expense, and reducing the risks of poor pharmaceutical inventory management seem to be the priorities. Of course, over the last twenty years, solutions have been proposed to curb this phenomenon as optimally as possible, but beyond the logistics and inventory management aspect, the question arises about dimensioning new warehouses with new services related to product management. This dimensioning process can take several months and lead to unfortunate incidents that can affect the quality of care, a luxury that organizations cannot afford. Determining dimensioning parameters implies knowledge of existing methods in terms of quantifying and estimating pharmaceutical needs, as well as the factors that influence these needs.

The objective of this paper has been to review these methods and the factors influencing the consumption of pharmaceutical products. Six quantification methods have emerged, three of which are the most widespread, such as consumption and morbidity methods, from the 1980s to the present day, proposed by academics as well as practitioners in the health sector who have recognized the potential of their applications. The application of these approaches is based on a set of preconditions. In this sense, the determination of consumption factors was essential. A set of 34 factors were therefore listed and grouped into three categories.

As this paper is the first in a series, the determination of these factors will allow future work to contextualize them in the French health system. The resulting factors should be studied in order to prove the effectiveness and correlations with the consumption of drugs and health products in general. Robust parameters for the dimensioning and management of stocks should result from this perspective. Moreover, the estimation methods found in the literature do not make use of Big Data techniques. It would be appropriate to propose tools for predicting the needs of hospital centers, both by functional unit and by health establishment, along the lines of artificial intelligence and big data techniques. Such are the prospects of our future work.

## 6 ACKNOWLEDGMENTS

This work has been partially supported by MIAI@Grenoble Alpes, (ANR-19-P3IA-0003).